\documentclass[twocolumn,showpacs,preprintnumbers,amsmath,amssymb]{revtex4}
\usepackage{tabularx,graphicx}

\usepackage{color}
\usepackage{hyperref}
\hypersetup{
    colorlinks=true,
    linkcolor=blue,
    filecolor=blue,      
    urlcolor=blue,
}

\begin{document}

\newcommand{\beq}{\begin{equation}}
\newcommand{\eeq}{\end{equation}}
\newcommand{\beqn}{\begin{eqnarray}}
\newcommand{\eeqn}{\end{eqnarray}}
\newcommand{\bmath}{\begin{subequations}}
\newcommand{\emath}{\end{subequations}}
\newcommand{\bra}[1]{\langle #1|}
\newcommand{\ket}[1]{|#1\rangle}

\title{Spinning superconductors and ferromagnets}
\author{J. E. Hirsch\footnote{$email$: jhirsch@ucsd.edu} }
\address{Department of Physics, University of California, San Diego,
La Jolla, CA 92093-0319}

\begin{abstract} 
When a magnetic field is applied to a ferromagnetic body it starts to spin (Einstein-de Haas effect). This demonstrates the intimate connection between the electron's magnetic moment $\mu_B=e\hbar/2m_ec$, associated with its spin angular momentum $S=\hbar/2$, and ferromagnetism. When a magnetic field is applied to a superconducting body it also starts to spin (gyromagnetic effect), and when a normal metal in a magnetic field becomes superconducting and expels the magnetic field (Meissner effect) the body also starts to spin. Yet according to the conventional theory of superconductivity the electron's spin only role is to label states, and the electron's magnetic moment plays no role in superconductivity. Instead, within the unconventional theory of hole superconductivity, the electron's spin and associated magnetic moment play a fundamental role in superconductivity. Just like in ferromagnets the magnetization of  superconductors is predicted to result from an aggregation of magnetic moments with angular momenta $\hbar/2$. This 
gives rise to a ``Spin Meissner effect'', the existence of a spin current in the ground state of superconductors. The theory explains how a superconducting body starts spinning when it expels magnetic fields, which we argue is not  explained by the conventional theory, it provides a dynamical explanation for the 
Meissner effect, which we argue the conventional theory cannot do, and it explains how
supercurrents stop without dissipation, which we argue the conventional theory fails to explain. Essential elements of the theory 
of hole superconductivity are that superconductivity is driven by lowering of kinetic energy, which we have also proposed
 is true for ferromagnets], that the normal state charge carriers in superconducting materials are holes, and that
 the spin-orbit interaction plays a key role in superconductivity. The theory is proposed to apply to all superconductors.
 \end{abstract}
\maketitle

\section{introduction}
The first paragraph of the  seminal paper of Einstein and de Haas \cite{edh} states: ``{\it When it had been discovered by Oersted that magnetic actions are exerted not only by permanent magnets, bnt also by electric currents, there seemed to be two entirely different ways in which a magnetic field can be produced. This conception, however, could hardly be considered as satisfactory and physicists soon tried to refer the two actions to one and the same cause. Ampere succeeded in doing so by his celebrated hypothesis of currents circulating around the molecules withont encountering any resistance.}''

Progress in physics often occurs through unification, where seemingly unrelated phenomena are recognized as manifestations of the same underlying physics. So it happened with the celebrated
experiment of Einstein and de Haas that proved the association between magnetic moment and angular momentum in ferromagnetic materials. 
While it may seem obvious to us now, it was not at all obvious at that time that the experiment would confirm the hoped-for unification. This is illustrated by the caveats expressed by Einstein and de Haas \cite{edh}: ``{\it It cannot be denied that these views call forth some objections. One of these is even more serious than it was in Ampere's days; it is difficult to conceive a circulation of electricity free from all resistance and therefore continuing forever. Indeed, according to
Maxwell's equations circulating electrons must lose their energy by radiation...The energy of the revolving
electrons would therefore be a true zero point energy. In the opinion of many physicists however, the existence of an energy of
this kind is very improbable. It appears by these remarks that after all as much may be said in favour of Ampere's hypothesis as against it }''. Their experiment of course proved that these caveats were
unfounded.

It is interesting to note that shortly  before Einstein and de Haas, Kammerlingh Onnes pointed out
a physical realization of Ampere's molecular currents, not in ferromagnets but in superconductors.
In his 1914 paper: ``{\it Further experiments with liquid helium. J. The imitation of an Ampere molecular
current or of a permanent magnet by means of a supraconductor}'' \cite{onnes} Onnes concluded that
``{\it it is possible
in a conductor without electromotive force or  leads from outside
to maintain a current permanently and thus to imitate  a permanent magnet or better a molecular
current as imagined by Ampere}''.

In this paper we argue that the commonalities between ferromagnets and superconductors go
much further than generally assumed.

\section{magnetic moment and angular momentum}

\begin{table}
\caption{Experiments on rotation and magnetization.}
\begin{tabular}{l | c | c | c  }
  \hline
    \hline
& Rotation by    &  Magnetization by   \cr
&   magnetization   &  rotation  \cr
 \hline  \hline
Ferromagnets & Einstein-de Haas & Barnett& \cr
&  effect (1915) \cite{edh} & effect (1915) \cite{barnett}& \cr
 \hline
Superconductors & Gyromagnetic  effect & London moment& \cr
  & (Kikoin and& (Becker et al (1933) \cite{becker}; & \cr
    & Gubar (1940)) \cite{gyro}   & Hildebrandt (1964)) \cite{hild}& \cr
 \hline
 \end{tabular}
\end{table}

The experiments showing the connection between angular momentum and magnetic moment
in ferromagnets and superconductors are summarized in Table I. The experimental
results follow from conservation laws. However the question of
$how$ angular momentum causes magnetization or magnetization causes angular
momentum is a separate question that needs to be addressed and understood.
For ferromagnets, only recently has this question started
to be considered   \cite{chud,niu}, and remains unsolved. For superconductors it isn't
even acknowledged that such a question exists and needs to be answered.

In this paper we address and solve this question  for superconductors. For ferromagnets
it remains an open question.

\section{magnetization in superconductors}
In ferromagnets, the magnetization is due to the intrinsic magnetic moment of the electron.
Assuming  there is no spin-orbit interaction, the Ampere molecular currents 
consist of each electron spinning around its axis, with mechanical angular momentum $S=\hbar/2$,
contributing magnetic moment 
\beq
\mu=g\frac{e}{2m_ec}S=\mu_B 
\eeq
with $g=2$ and  $\mu_B$ the Bohr magneton.

 In superconductors, the supercurrent resides within a London penetration depth ($\lambda_L$) of the 
surface. Consider a long cylinder in a magnetic field $H$ along its axis. The magnetization of the
cylinder is $M=H/4\pi$ so that the magnetic field in the interior is zero. With 
$n_s=$ number of superfluid electrons per unit volume, the magnetic moment $\mu$ contributed by each electron is
\beq 
\mu=\frac{M}{n_s}=\frac{H}{4\pi n_s}=\frac{e}{2m_e c}\ell
\eeq
since $g=1$ for orbital motion. Here, $\ell$ is the mechanical angular momentum contributed by each superfluid
electron. The mechanical momentum of electrons in the Meissner current is 
\beq
m_ev=\frac{e}{c}A=\frac{e}{c}\lambda_LH .
\eeq
Using the well-known relation between London penetration depth and superfluid density \cite{tinkham}
\beq
\frac{1}{\lambda_L^2}=\frac{4\pi n_se^2}{m_ec^2}
\eeq
it follows from Eqs. (2), (3) and (4) that
\beq
\ell=m_ev(2\lambda_L) .
\eeq
Therefore the Meissner magnetization $M=H/(4\pi)$ can be understood as arising from each superfluid electron in an orbit
of radius $2\lambda_L$ with orbital angular momentum $\ell$ given by Eq. (5) contributing its orbital magnetic moment Eq. (2).
These orbital motions cancel out in the interior of the cylinder but not within a London penetration depth of the
surface, giving rise to the surface Meissner current. 

As we will show in what follows, these Amperian `molecular currents' exist in superconductors also in the absence
of an applied magnetic field, with each electron carrying orbital angular momentum $\ell=\hbar/2$ \cite{sm,bohr}. 
Both in the superconductor and the ferromagnet the `molecular currents' in the interior cancel out leaving
a surface current per unit length $cM$ to give rise to the macroscopic magnetization. This is shown
schematically in Fig. 1. For superconductors, 
the radius of these currents is $2\lambda_L$ and the speed of motion is $v_\sigma=\hbar/(4m_e\lambda_L)$,
while for ferromagnets  the radius is $r_q=\hbar/(2m_ec)$ and the speed of motion is $c$.
In both cases the angular momentum is $\hbar/2$.  This shows that a remarkable connection exists between superconductors and ferromagnets that was heretofore unrecognized.

           \begin{figure}
 \resizebox{8.5cm}{!}{\includegraphics[width=6cm]{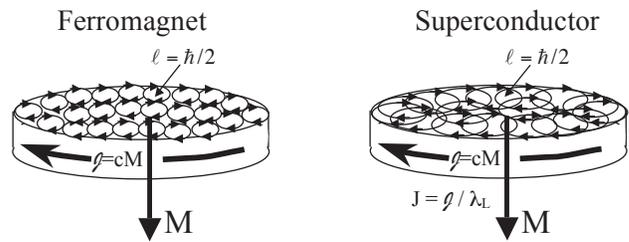}}
 \caption { Common physics of superconductors and ferromagnets. The mechanical angular momentum of the
 carriers giving rise to magnetization is $\hbar/2$ in both cases.
 }
 \label{figure1}
 \end{figure} 
 
In the presence of spin-orbit interaction, the canonical momentum of an electron of spin $\sigma$ is given by
\beq
\vec{p}=m_e\vec{v}_\sigma+\frac{e}{c}(\vec{A}+\vec{A}_\sigma)
\eeq
where
\beq
\vec{A}_\sigma=\frac{\hbar}{4m_ec}\vec{\sigma}\times\vec{E}
\eeq
is the spin-orbit vector potential derived from the Dirac equation in the presence of  electric field $\vec{E}$ \cite{bohr}.
Electrons in the superfluid move in a positive background of charge density $|e|n_s$, which gives rise to a radial
electric field (in a cylindrical geometry)
\beq
\vec{E}=2\pi |e|n_s\vec{r}
\eeq
so that
\beq
\vec{A}_\sigma=-2\pi n_s\frac{e\hbar}{2m_ec}\frac{\vec{\sigma}\times\vec{r}}{2}\equiv \frac{\vec{H}_\sigma\times\vec{r}}{2}
\eeq
with
\beq
\vec{H}_\sigma=-2\pi n_s\mu_B\vec{\sigma}=2\pi n_s\vec{\mu}=\frac{\hbar c}{4e\lambda_L^2} \vec{\sigma}
\eeq
the `spin-orbit magnetic field'. By this we mean, $H_\sigma$ is the magnetic field that would exert the same force
on the moving charge e as the electric field exerts on the moving magnetic moment $\mu_B$.  

In the superconducting state the canonical momentum $p=0$, hence from Eq. (6)
\beq
m_e\vec{v}_\sigma=-\frac{e}{c}(\vec{A}+\vec{A}_\sigma)
\eeq
and the mechanical momentum is
\beq
m_ev_\sigma=\frac{e}{c}\lambda_L(H+H_\sigma) ;
\eeq
In particular, for $H=0$ Eqs. (10) and (12) yield, using Eq. (4)
\beq
m_ev_\sigma=\frac{1}{2\lambda_L}\frac{\hbar}{2}
\eeq
hence from Eq. (5)
\beq
\ell=\frac{\hbar}{2}
\eeq

Therefore,   in the absence of applied magnetic field, superconducting electrons reside in
orbits of radius $2\lambda_L$, and carry orbital angular momentum $\hbar/2$, in opposite direction for
opposite spin. This gives rise to macroscopic zero point motion, a spin current that flows within a London
penetration depth of the surface, an Amperian `molecular current' without magnetization. As a magnetic field is applied, magnetization develops by having the electrons with spin
antiparallel to the magnetic field speed up and those with spin parallel to the magnetic field slow down,
giving rise to a charge current flowing within $\lambda_L$ of the surface. The lower critical field
of type II superconductors
\beq
H_{c1}=-\frac{\hbar c}{4e\lambda_L^2}
\eeq
is precisely the same as the spin-orbit field Eq. 10. This implies that when the applied
magnetic field reaches this value one of the spin current components becomes zero, the
supercurrent stops flowing and superconductivity is destroyed, allowing the magnetic
 field to penetrate \cite{sm,copses}.

In recent experimental work \cite{tallon,tallon2}  it has been shown that the critical current that gives rise to onset of dissipation
in type II superconductors is   given by the universal form
\beq
J_c=\frac{c}{4\pi\lambda_L}H_{c1}
\eeq
with $H_{c1}$ given by Eq. (15), independent of what the actual value of the magnetic field is at the boundary 
of the sample. The actual value of the magnetic field can vary by orders of magnitude depending on the sample
size and geometry  \cite{tallon,tallon2}. This finding implies that what determines the onset of dissipation is not vortex formation, but 
rather that the carriers of the supercurrent $J_c=n_sev_c$ reach a critical speed $v_c$. From Eqs. (15) and (16)
\beq
v_c=\frac{\hbar}{4m_e\lambda_L}
\eeq
which is the speed of the spin current Eq. (13). 
In other words, dissipation sets in when one of the components
of the spin current comes to a stop. At this point the extra kinetic energy due to the charge current flow becomes equal to the condensation energy \cite{copses}.

            \begin{figure}
 \resizebox{7.5cm}{!}{\includegraphics[width=6cm]{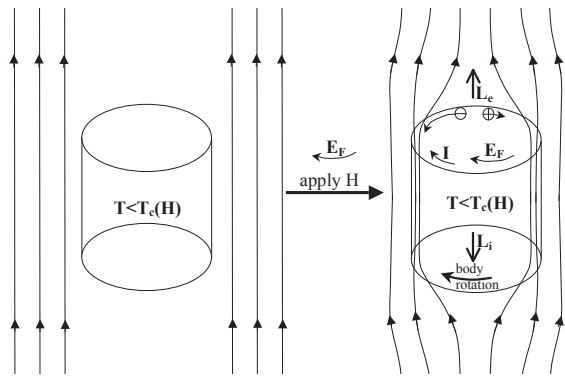}}
 \caption {  A magnetic field is applied to a superconductor at rest. A Faraday electric field $E_F$  is generated in the clockwise direction, 
 opposing the change in magnetic flux. $E_F$  pushes positive ions (negative electrons) in clockwise (counterclockwise) direction. The body acquires angular momentum
 $\vec{L}_i$  antiparallel to the
 applied magnetic field  and the supercurrent acquires angular momentum $\vec{L}_e=-\vec{L}_i$ parallel to the magnetic field.
 }
 \label{figure1}
 \end{figure}

\section{momentum transfer to the body}

When the magnetization of the superconducting body changes, the mechanical angular momentum of the body has to change
to compensate the mechanical angular momentum change of the electrons carrying the supercurrent.
It is easy to understand how this happens when an external magnetic field is applied to a superconductor \cite{momentum}:
the Faraday electric field generated by the change in magnetic flux exerts a force on the electrons in one direction
to generate the supercurrent and associated magnetization, and on  the positive ions in the opposite direction making the body as a whole
rotate, as shown in Figure 2. This has been verified experimentally (gyromagnetic effect in superconductors) \cite{gyro}.

Instead, it is not easy to understand how the angular momentum of the body changes when the supercurrent
and magnetization
change as a result of a change in temperature. In an applied magnetic field $H$,
there is a critical temperature $T_c(H)$. If initially $T$ is slightly below $T_c(H)$ a supercurrent flows, and
as the temperature is raised above $T_c(H)$ the supercurrent stops as the system becomes normal.
Conversely, starting with temperature slightly above $T_c(H)$, if the temperature is lowered a supercurrent is
spontaneously generated that expels the magnetic field as the system becomes superconducting.

The reason this is not easy to understand is because as these changes occur a Faraday electric field is generated
that opposes these changes. In other words, the Faraday field pushes both the electrons and the body to move
in direction exactly opposite to the direction they move. What is the driving force for the electron motion, and what
is the driving force for the body motion?

The answers within the conventional theory of superconductivity 
\cite{eilen,scal,halp} are shown schematically in Figure 3. 
The conventional theory assumes that when the system becomes normal
Cooper pairs carrying the supercurrent dissociate, the resulting normal
electrons inherit the center of mass momentum of the Cooper pairs, and transfer it to the body as a whole through impurity scattering, as shown
in Fig. 3a. In the reverse transformation (Meissner effect), it assumes that carriers becoming superconducting `spontaneously'
acquire the momentum of the supercurrent, leaving behind normal carriers with opposite momentum that
transfer it to the body as a whole through impurity scattering, as shown in Fig. 3b.

           \begin{figure}
 \resizebox{8.5cm}{!}{\includegraphics[width=6cm]{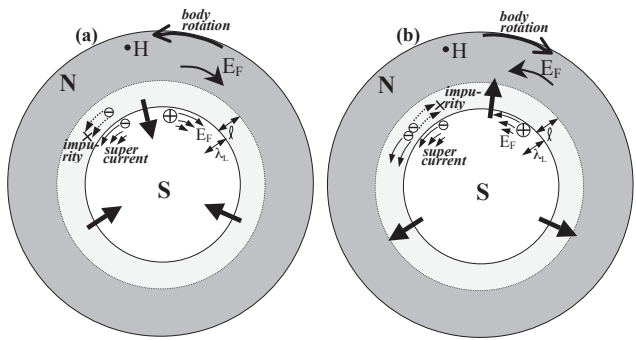}}
 \caption { Long cylinders seen from the top. Magnetic field $H$ is parallel to the cylinder axis.  (a) Superconductor goes normal
 (S-N transition). As the phase boundary moves inward, a clockwise Faraday
 electric field $E_F$ is generated. $E_F$ pushes positive ions in clockwise direction, and negative electrons in the supercurrent
 in counterclockwise direction. Instead, body starts rotating in counterclockwise direction and the counterclockwise 
 supercurrent stops.
  According to the conventional theory, the supercurrent is stopped by impurity scattering, and these scattering processes
 transfer the counterclockwise momentum of the supercurrent to the body.
 (b) Reverse (N-S, or Meissner) transition.  As the phase boundary moves outward, a counterclockwise Faraday
 electric field $E_F$ is generated. $E_F$ pushes positive ions in counterclockwise direction, and negative electrons in the supercurrent
 in clockwise direction. Instead, body starts rotating in clockwise direction and a counterclockwise 
 supercurrent is spontaneously generated.
 }
 \label{figure1}
 \end{figure} 
 
 Note  the role played by the Faraday field. In the S-N transition, it accelerates the supercurrent in the S region close to the boundary and it
 transfers momentum to the ions in that region in clockwise direction, opposite to the motion of the body. Thus, the total  momentum that needs to be
 transferred to the body by impurity scattering is much larger (by a factor $R/(3\lambda_L)$, with $R$ the radius of the cylinder)
  than the momentum acquired by the body, since it has to compensate
 the momentum in opposite direction transferred by the Faraday field. Similarly in the N-S transition, with the signs of momenta
 reversed.
 
A fundamental   problem with the conventional explanation   is that it relies on impurity scattering to transfer momentum between the electrons and the
body as a whole. Impurity scattering is an inherently  irreversible process, whether it is elastic or inelastic.
A normal electron with the momentum of a supercurrent electron will scatter in a random direction, but a normal electron incident from a 
random direction will not be scattered to acquire the momentum of the supercurrent electron.  
These scattering processes increase the entropy of the universe, no matter how slowly the transition proceeds. However, 
the superconductor-normal transition is known to be a thermodynamically reversible process, that can occur
(under ideal conditions) with no change in the entropy of the universe \cite{gorter,revers}. A valid theory of superconductivity has to have
the ability to explain how momentum is transferred {\it in a reversible way} between the electronic degrees of freedom
and the body as a whole. We have called this `the central question in superconductivity' \cite{centralquestion}.

             \begin{figure}
 \resizebox{7.5cm}{!}{\includegraphics[width=6cm]{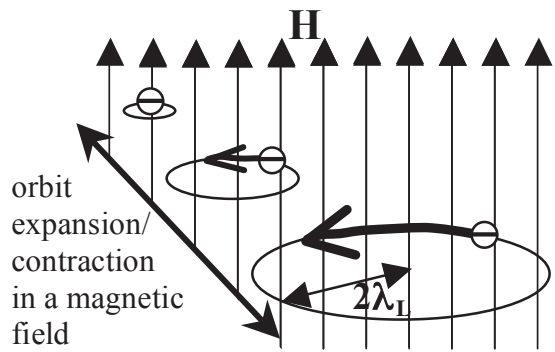}}
 \caption { When an electron expands or contracts its orbit in a perpendicular magnetic field its azimuthal
 velocity changes proportionally to the radius of the orbit due to the azimuthal Lorentz force acting on the radially 
 outgoing or ingoing charge.
 }
 \label{figure1}
 \end{figure}
 
 The theory of hole superconductivity \cite{holesc} provides a possible answer to this question. We are not aware that any other answer
exists or is possible.  There are two key parts to the question: (1) how do the electrons lose or acquire their momentum? (2) how does
the body acquire or lose its momentum? The essence of the answer is that the momentum transfer is mediated by
the electromagnetic field \cite{momentum}. It does not involve any scattering processes, hence it is inherently $reversible$.

First, how do electrons acquire or lose their momentum reversibly? We   propose that this results from expansion or contraction
of electronic orbits, as shown schematically in Fig. 4. 
The expansion of the orbits when the system goes superconducting is driven by lowering of quantum
kinetic energy \cite{meissnerorigin}. When the orbit expands from a microscopic radius to an orbit of radius $r$, the magnetic Lorentz force on the radially outgoing  motion generates an azimuthal velocity \cite{copses}
\beq
v_\phi=-\frac{er}{2m_ec}H  .
\eeq
Thus, when the orbit expands to radius $r=2\lambda_L$ the electron acquires the azimuthal speed of the Meissner current
Eq. (3). Similarly, when the orbit shrinks from radius $2\lambda_L$ to a microscopic radius the azimuthal Lorentz
force acts in the opposite direction and the supercurrent stops. According to the theory of hole superconductivity when electrons
form a Cooper pair and become part of the superconducting condensate their orbits expand from microscopic radius to radius $2\lambda_L$, and conversely
when electrons depair and become normal the orbits shrink from radius $2\lambda_L$ to a microscopic radius.
This reversible expansion or contraction of the orbits explains how the supercurrent starts and stops in a given
external magnetic field.

The second part of the question is, how does the body as a whole acquire a compensating momentum in the opposite direction?
In the  process just described, momentum conservation holds because  as the orbits expand or contract and
the azimuthal momentum of the electron changes
a compensating azimuthal momentum is stored in the electromagnetic field, as explained in \cite{momentum}. That momentum
is retrieved by a radial flow of normal charge to compensate for the radial charge redistribution that occurs when the
electronic orbit expands or contracts. Here is where the hole-like nature of the normal state charge carriers plays a key role.

             \begin{figure}
 \resizebox{8.5cm}{!}{\includegraphics[width=6cm]{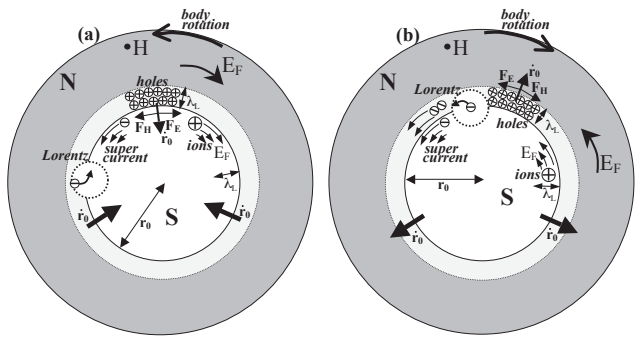}}
 \caption { (a) S-N transition and (b)  N-S transition according to the theory of hole superconductivity.
The supercurrent electrons acquire or lose their momentum through the Lorentz force as the orbits
expand or shrink (dotted circles). The expansion or contraction of the orbits causes radial charge flow,
and there is a radial backflow of normal state charge to preserve charge neutrality.
The backflow is in the form of normal holes moving in the same direction as the phase boundary. Electric and magnetic
forces on the holes ($F_E$ and $F_H$) are balanced so the motion is purely radial. }
 \label{figure1}
 \end{figure}

Figure 5 explains the physics of the process. In (a), the superconducting region is shrinking. $2\lambda_L$ orbits 
at the boundary of the N-S region are contracting, which imparts the electron with a clockwise azimuthal impulse,
stopping the supercurrent. At the same time, to compensate for the radial inflow of negative charge, a backflow of 
normal charge occurs: normal state hole carriers flow radially inward at speed $\dot{r}_0$, the speed of motion of the phase boundary.  Similarly, when the superconducting region expands (Fig. 5(b)), expansion of the orbits imparts electrons
with a counterclockwise azimuthal impulse, generating the supercurrent, there is a radial outflow of negative
charge and backflow occurs as a compensating radial outflow of positive hole carriers to maintain charge neutrality.

              \begin{figure}
 \resizebox{8.5cm}{!}{\includegraphics[width=6cm]{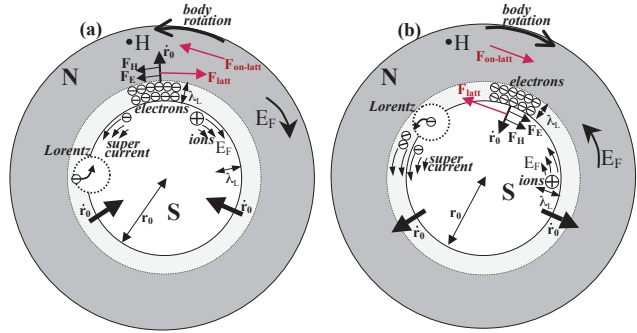}}
 \caption { Figure 5  redrawn replacing the backflowing holes by electrons flowing in the opposite direction. 
 The electric and magnetic forces on electrons
 $F_E$ and $F_H$ point in the same direction, making it clear that another force must exist,
 $F_{latt}$, exerted by the periodic potential of the ions on the charge carriers.
 By Newton's third law, an equal and opposite force is exerted by the charge carriers on the ions,
 $F_{on-latt}$, that makes the body rotate.
 }
 \label{figure1}
 \end{figure}

The azimuthal forces on this backflow of inflowing or outflowing hole carriers 
are $F_H$ and $F_E$ shown in Fig. 5. $F_H$ is the magnetic Lorentz force  and 
$F_E$ is the electric force due to the Faraday field $E_F$, which  is given according to Faraday's law  by
\beq
E_F=\frac{\dot{r}_0}{c}H .
\eeq
Thus, $eE_F$ is precisely the magnetic Lorentz force on the hole carriers   in opposite direction to the electric force, the 
azimuthal forces on the holes exactly cancel out and the hole motion is radial.

Now in order to understand how the transfer of momentum happens we need to remember that holes are just a theoretical construct, the actual carriers of charge in a metal are always electrons of negative charge.
So we redraw the backflows of holes in Figure 5 in terms of electrons that flow in opposite direction to the
motion of the phase boundary in Figure 6. Note that now the electric and magnetic forces on these electrons are $not$ balanced, rather
they point in the same direction. To recover force balance we need to
add another force $F_{latt}$, which is a transverse force that the periodic ionic potential exerts on 
the charge carriers moving in crossed electric and magnetic field {\it when the charge carriers are holes}.
By Newton's third law, there is an equal opposite force exerted by the electrons on the ions, which
we denote by $F_{on-latt}$. This force acts in the direction of the body rotation, it is in fact the force that makes the body rotate.

If the normal state charge carriers were electrons rather than holes, there would be no $F_{latt}$ in Fig. 6 and the transverse
electric and magnetic forces would be unbalanced resulting in an azimuthal charge flow that would give rise to 
dissipation and irreversibility. Note also that in the absence of backflow there would be an azimuthal current induced by 
the Faraday field that would also give rise to dissipation and irreversibility \cite{revers}.

In summary, the transverse force exerted by the ions on the backflowing electrons
{\it for materials where the normal state charge carriers are holes} transfers momentum to the
electrons, and by the same token electrons transfer momentum to the ions and hence to the body as a whole,
making the body rotate and accounting for momentum conservation. This explains how the momentum
acquired or lost by the supercurrent becomes momentum of the body as a whole without any irreversible
scattering processes. The essential ingredient to make this happen is that the normal state carriers
are holes, or equivalently have negative effective mass. 
For this reason superconductivity can only
occur if normal state charge carriers are holes. If normal state carriers are electrons there is no 
reversible mechanism for momentum transfer between charge carriers and the body and 
the normal-superconductor and superconductor-normal transitions, as we know them to take place in 
nature, cannot take place.

Here we have only discussed the simplest scenario where the superconducting phase grows or shrinks 
with azimuthal symmetry, as shown in Figs. 3, 5 and 6. The same principles are of course at work  in more general situations,
e.g. where the superconducting phase grows by nucleation of many different domains that develop separately and 
then merge. Growing domains expel negative charge as they grow, and
there is a backflow of negative charge carried by outflowing normal hole-like carriers. It has just come to our attention that
some elements of this physics have been proposed earlier  by W. H. Cherry and J. I. Gittleman \cite{cg} as being relevant to
the understanding of the Meissner effect.

\section{conclusions}
Since the work of Einstein and de Haas we have known that angular momentum and magnetic moment
are intimately coupled. Both in ferromagnets and in superconductors angular momentum and magnetic moments
are part of the essential physics. In ferromagnets, we know  that
the electron intrinsic magnetic moment $\mu_B$ and its  spin angular momentum
$\hbar/2$ play a key role. However, within the conventional theory of superconductivity \cite{tinkham} the electron spin plays only the
role of labeling states, and the electron's intrinsic magnetic moment plays no role.
Instead, within the theory of hole superconductivity the electron spin and its associated magnetic moment play a
key role.  The theory predicts that just like in ferromagnets there are Amperian ``molecular currents'' composed of elementary units
with mechanical angular momentum $\hbar/2$. These elementary units are electrons in $2\lambda_L$ mesoscopic orbits
orbiting with speed Eq. (17). They acquire this speed as the orbit expands, through the spin-orbit interaction of the
electron magnetic moment with the background electric field of the positive ions \cite{sm,bohr}. 
The $orbital$ magnetic moment of electrons in these orbits is $\mu_B/2$, in direction opposite to the 
intrinsic electron magnetic moment \cite{100}. When an external magnetic field is applied, the speed of electrons in these orbits is
modified to give rise to the macroscopic magnetization that opposes flux penetration. If the system becomes superconducting
in a magnetic field, the expansion of the orbits provides a dynamical explanation of the Meissner effect, that is
not provided by the conventional theory.

Using these concepts together with the essential fact that the normal state charge carriers are holes we are able to
completely describe the physics that takes place when either magnetization or angular momentum \cite{lm,lm2} changes in 
superconductors,
in a way that satisfies the conservation laws, explains the physical processes by which  momentum is transferred between different components
of the system,  and respects the fact that the processes are reversible under ideal
conditions. We cannot say the same thing about ferromagnets. Even though conservation laws will tell us what
happens when magnetization or angular momentum change in ferromagnets, we don't know yet  the detailed
processes by which change in one leads to change in the other. Perhaps thinking about  the way superconductors do it will
help to eventually understand  how ferromagnets do it.

\end{document}